\documentclass[11pt,a4paper]{article}
\pdfoutput=1
\usepackage{jheppub}
\usepackage[utf8]{inputenc}

\usepackage{color}
\usepackage{amsmath}
\usepackage{pifont}
\usepackage{bbold}

\usepackage{bbm}
\usepackage{verbatim}   
\usepackage{subfigure}
\usepackage{acronym}

\usepackage{amsfonts}
\usepackage{amssymb}
\usepackage{mathtools}
\usepackage{mathrsfs}
\usepackage{graphicx}
\usepackage{multirow}
\usepackage{slashed}
\usepackage{array} 
\newcolumntype{C}{>{\hfil$}p{14pt}<{$\hfil}}

\usepackage{graphicx}
\usepackage{empheq}
\usepackage{caption}
\usepackage{tikz}
\usetikzlibrary{decorations.pathreplacing}

\newcommand{\diff}[2]{\frac{\mathrm{d}#1 }{\mathrm{d}#2 }}
\newcommand{\ddiff}[2]{\frac{\mathrm{d}^2 #1 }{\mathrm{d} #2^2 }}
\newcommand{\pd}[2]{\frac{\partial #1}{\partial #2}}
\renewcommand{\Im}{\mathrm{Im}\,}

\title{Tunnelling of Charged Particles from Black Holes}

\author[1]{George Johnson}
\emailAdd{george.johnson@physics.ox.ac.uk}

\affiliation[1]{Rudolf Peierls Centre for Theoretical Physics, University of Oxford, OX1 3NP, United Kingdom}

\abstract{We state and explain the total rate at which a charged non-rotating black hole emits charged particles, taking into account both Hawking radiation and Schwinger pair production simultaneously. We give concrete formulae for this emission rate in certain limits, with the greatest simplification occurring when the black hole is much larger than the particle's Compton wavelength. We provide an interpretation of the result in terms of a tunnelling process, both through the black hole horizon and the surrounding electric field, and comment on how suppression due to tunnelling modifies the emission spectrum.}

\begin{document}

\maketitle

\section{Introduction}

In 1974, Hawking \cite{hawking1} demonstrated the surprising result that, contrary to classical expectations, black holes could emit particles, and that they did so with a precisely thermal spectrum. Particle creation in another context, namely that by a static \textit{electric} field, had been understood a quarter of a century earlier, with the work of Schwinger \cite{schwinger}. Since it is possible for a black hole to be electrically charged, we expect that black holes can lose mass and charge through this process also.

Whilst there has been a wealth of research into the spectrum of radiation of uncharged particles from black holes, for which Schwinger pair production is irrelevant, and into the nature of Schwinger pair production outside a charged black hole, with thermal effects ignored, there has not been such detailed analysis of the interplay of the two effects --- of the entirety of emission from a hot electrically charged black hole.  Where both processes are considered, it is usually as applied to \textit{different} species of particle --- that is, one considers Hawking radiation of photons, say, and Schwinger production of electrons.

In this paper, we clarify how the total rate at which a charged black hole emits a particular species of charged particle is determined by both production processes. Not only is it interesting in its own right to understand the nature of radiation in the general case, it is important in understanding how black hole decay behaves in certain theoretically interesting limits. The weak gravity conjecture \cite{wgc}, for instance, which loosely speaking states that $q > m$ for some particle in the spectrum of any consistent quantum theory of gravity, is motivated in large part by arguments involving the decay of extremal black holes. We thus wish to understand precisely how black hole radiation behaves in the limit that the black hole charge, or indeed the emitted particle's charge, tends to its mass.

This paper is divided into five sections. In Section 2, we discuss the general theory of Hawking radiation of uncharged particles and of Schwinger pair production in flat spacetime. In Section 3, we take account of both of these phenomena to give an exact formula for the rate at which charged black holes lose energy (or indeed some other quantity), in terms of transmission coefficients that can be calculated, at least in principle, by solving a differential equation. In Section 4 we provide approximate formulae for these transmission coefficients in terms of tunnelling integrals, as well as an interpretation of black hole decay as a two-stage tunnelling process. Finally, in Section 5, we summarise our results.

We will use the terms `boson' and `fermion' to mean spin-0 and spin-1/2 particles respectively; we will not consider particles of spin $s \geq 1$ in this paper. We use the $\begin{matrix} -&+&+&+\end{matrix}$ metric convention, and take $c = \hbar = 4 \pi \epsilon_0 = G = k_B = 1$ throughout.

\section{Preliminary Theory}

\subsection{Energetics of a charged black hole}
\label{firstlaw}

A charged non-rotating black hole is described by the Reissner-Nordstr\"{o}m metric:
\begin{equation}
\mathrm{d}s^2 = -\left(1-\frac{2M}{r}+\frac{Q^2}{r^2}\right)\mathrm{d}t^2 + \left(1-\frac{2M}{r}+\frac{Q^2}{r^2}\right)^{-1} \mathrm{d}r^2 + r^2 \mathrm{d}\Omega^2 \,,
\end{equation}
where $Q$ is the charge of the black hole and $M$ is its total (ADM) energy. There is an event horizon at $r_+ = M + \sqrt{M^2-Q^2}$ and a singularity at $r=0$. Throughout this paper we will write $f(r) = g_{tt}$. The electromagnetic potential outside the black hole is
\begin{equation}
A = -\frac{Q}{r} \mathrm{d}t \,.
\end{equation}

We can consider the energy $M$ of a charged black hole to have two contributions: one, from the mass energy stored inside the hole itself, $M_\mathrm{irr}$, and two, from the energy stored in the electric field outside the black hole, $U$:
\begin{align}
M_\mathrm{irr} &\coloneqq\frac{r_+}{2} = \frac{1}{2}M + \frac{1}{2}\sqrt{M^2-Q^2}\,,\\
U &\coloneqq \frac{1}{8 \pi} \int E^2 \, \mathrm{d}V = \frac{Q^2}{2 r_+} = \frac{Q^2}{2M + 2\sqrt{M^2-Q^2}} \,.
\end{align}
One can check that $M_\mathrm{irr} + U = M$. Since the area of the black hole is a monotonic function of its irreducible mass, the classical area theorem $\mathrm{d} A \geq 0$ corresponds to
\begin{equation}
\mathrm{d}M_\mathrm{irr} \geq 0 \,.
\end{equation}
In other words, we can classically extract energy only from the external field of the black hole, and never from its irreducible mass. By extracting energy in such a way that $\mathrm{d}M_\mathrm{irr} = 0$, we find that the maximum energy that can be extracted from a charged black hole is $U$, after which point the charge is reduced to zero. The field energy $U$ can constitute up to half of the total energy of the black hole. 

If a black hole emits a particle of charge $q$ and energy $\omega$, the first law of black hole mechanics reads
\begin{align}
\label{1law}
\mathrm{d}M &= \frac{\kappa}{8 \pi}\mathrm{d}A + \Phi \mathrm{d}Q \,, \\
\label{1law2}
-\omega &= \frac{\sqrt{M^2-Q^2}}{8 \pi r_+^2}\mathrm{d}A - \frac{qQ}{r_+} \,,
\end{align}
where $\kappa$ is the surface gravity and $\Phi = Q/r_+$ is the electric potential at the horizon. We see that emission is divided into two qualitatively different regimes: one, where $\omega \leq q\Phi$, which is classically allowed ($\mathrm{d}A \geq0$), and one,  $\omega > q\Phi$, which is not, ($\mathrm{d}A <0$). In fact particles with energies greater than $q Q/r_+$ \textit{can} be radiated when quantum effects are taken into consideration; this is precisely Hawking radiation.

Note that one needs to take care with this equation when the black hole is extremal, that is, when $M = Q$. Then the surface gravity vanishes and Eq. \eqref{1law2} appears to read
\begin{equation}
\omega = q \,.
\end{equation}
That is, the black hole can only emit particles of energy $q$. This is false, on account that $A$ is not a differentiable function of $M,Q$ at $M=Q$. Informally, $\mathrm{d}A$ becomes infinite for small perturbations about extremality.

In this case, we know that $\omega$ cannot be larger than $q$, for this would result in a super-extremal black hole and an associated naked singularity. A more careful calculation shows that only  for $\omega < q$ does the area increase in the emission process, and hence just as for non-extremal black holes, $\omega \leq q$ corresponds to the classically allowed range of emission energies.

\subsection{Hawking radiation of uncharged particles}
\label{hawkunch}

Classically, it is not possible for a particle to emerge from behind the event horizon. However, the surprising result found by Hawking \cite{hawking1} is that the collapse of matter to form a black hole results in the emission of radiation that persists at late time. The fact that any particles are produced at all can be attributed to the time-dependent nature of the spacetime during collapse. The spectrum of radiation is precisely that which one would expect from a thermodynamic black body at a temperature given by $T_\mathrm{BH} = \kappa/2\pi$. 

For a Reissner-Nordstr\"{o}m black hole this temperature is
\begin{equation}
\label{temp}
T_\mathrm{BH} = \frac{\sqrt{M^2-Q^2}}{2 \pi r_+^2} \,,
\end{equation}
and the result of Hawking is that the mean number of particles produced in the mode $\omega$ is
\begin{equation}
n(\omega) = \frac{1}{\exp(\omega/T_\mathrm{BH}) \mp 1} \,,
\end{equation}
where the upper sign refers to bosons and the lower sign to fermions. To convert this into the total rate at which the black hole loses energy, we need to integrate over all momentum modes, and multiply by an absorption cross-section, or grey-body factor, $\sigma(\omega)$. This describes how the produced particles escape to infinity; if flux is reflected back into the black hole by the external gravitational field, the greybody factor will be correspondingly smaller. The grey-body factor is typically calculated in a time-reversed setup, whereby the flux at the horizon is purely \textit{ingoing}. Thus the total rate of energy loss is

\begin{equation}
\label{stefan}
-\diff{M}{t} = g \int \frac{\mathrm{d}^3 k}{(2 \pi)^3} \,\omega n(\omega) \sigma(\omega)= \frac{g}{2 \pi^2}\int_m^\infty \mathrm{d}\omega \, \frac{\omega^2 k }{\exp(\omega/T_\mathrm{BH}) \mp 1}\sigma(\omega) \,,
\end{equation}
where $g$ is the number of degrees of freedom of the particle and $k$ is its momentum, $k = \sqrt{\omega^2 - m^2}$. If the interaction with the black hole exterior were negligible, the absorption cross-section would simply be the cross-sectional area of the black hole, $A/4$, and one can show that Eq. \eqref{stefan} reduces to the Stefan-Boltzmann law. The rate at which the black hole loses some other quantity, such as charge, can be found by replacing one factor of $\omega$ inside the integral by the quantity of interest.

For Hawking radiation from a Schwarzschild black hole, this grey-body factor depends weakly on $\omega$ \cite{unruh}, and the exponential dependence of the Planck distribution has the largest impact on the emission spectrum. However, as we will find, for charged particles the grey-body factor can also be exponentially sensitive to $\omega$, resulting in large suppression of radiation in parts of the emission spectrum.

\begin{figure}[h]
	\centering
	\includegraphics[width=\textwidth]{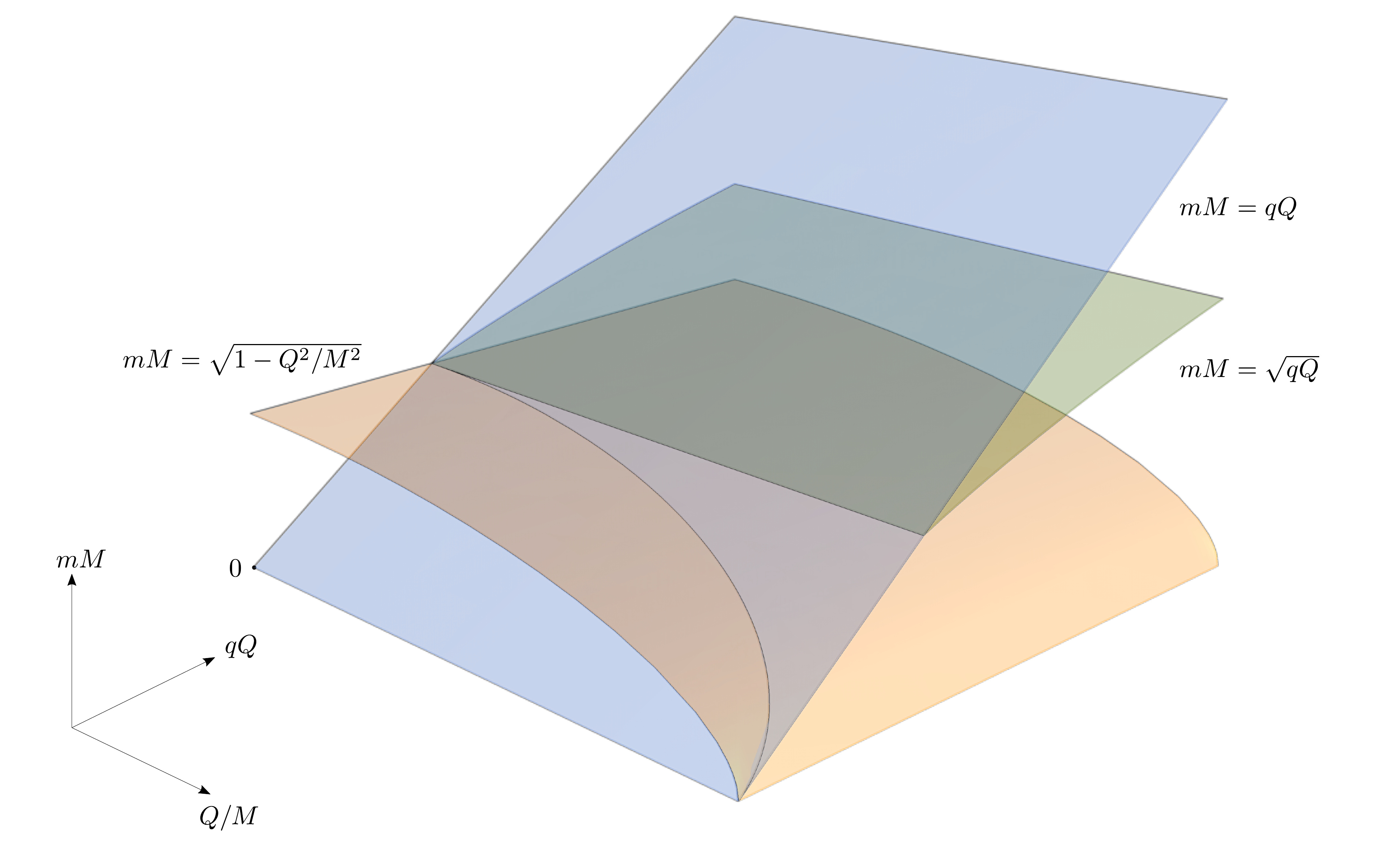}
	\caption{A plot of the regions in parameter space in which the two emission processes are important. On the three axes are plotted the mass of the emitted particle, the charge of the emitted particle, and the mass-to-charge ratio of the black hole. Beneath the orange surface, Hawking radiation is appreciable. Beneath the blue plane, Schwinger pair production is possible. Beneath the green surface, Schwinger pair production is appreciable. Here $Q/M$ ranges between zero and one, whilst the other two quantities range between zero and three.}
	\label{bounds}
\end{figure}

In Fig. \ref{bounds} we give a crude plot of the regions in parameter space in which we expect Hawking radiation and Schwinger pair production to be important. In particular, we expect Hawking radiation to become important when the temperature of the black hole exceeds the mass of the particle, and the Schwinger effect to become important when the electrostatic force on the particle at the horizon exceeds its mass squared. Order one factors are ignored. See \cite{gibbons2} for a two-dimensional slice of this diagram.

\subsection{Schwinger production in flat spacetime}

It was shown by Schwinger \cite{schwinger} that a static electric field configuration is quantum mechanically unstable towards decay into a pair of charged particles. Energetically, the sum of the kinetic energies of the two particles must equal the change in potential energy between their positions, less the change in field energy stored in the electric field:
\begin{equation}
\label{energetics}
 q \Delta \Phi - \Delta U = \omega_1 + \omega_2 \,.
\end{equation}
In the case that the electric field remains unchanged, this puts a bound on the strength of the electric field required for the process to be energetically possible:
\begin{equation}
q \Delta \Phi > 2m \,.
\end{equation}
To understand the rate at which this process happens, we need to examine the field equations governing the behaviour of charged matter in an electric field. We will restrict attention to 1+1 dimensions for simplicity.

 In flat spacetime, the equation governing the behaviour of spin-0 charged particles in a static electric field $E(x)$ is the Klein-Gordon equation:
\begin{equation}
\label{kg}
-D^\mu D_\mu \phi + m^2\phi = \left[ \left(\pd{}{t} - iq A_t \right)^2 - \frac{\partial^2}{\partial x^2}  + m^2  \right] \phi = 0 \,,
\end{equation}
where $\partial_x A_t = E(x)$ (so $-q A_t$ is the  potential energy). For spin-1/2 particles, the relevant equation is the Dirac equation:
\begin{equation}
\label{dirac}
\slashed{D}\psi -m\psi = \left[ \gamma^0 \left(\pd{}{t} - iq A_t \right) + \gamma^1 \pd{}{x}  - m  \right] \psi = 0 \,.
\end{equation}
We can square this to give
\begin{equation}
\label{diracsq}
-\slashed{D}\slashed{D}\psi + m^2\psi = \left[\left(\pd{}{t} - iq A_t \right)^2  - \frac{\partial^2}{\partial x^2}  + m^2  + q\sigma^{\mu \nu} F_{\mu \nu}\right] \psi = 0 \,.
\end{equation}
We can always choose $\sigma^{01}$ to be diagonal, with diagonal elements $\pm i/2$, so that it measures spin along the $x$-axis. We thus find that each component $\psi_i$ of the fermion obeys an equation very similar to the bosonic equation, but with an additional imaginary term representing the coupling of the spin to the electric field:
\begin{equation}
\label{kgdirac}
\left[\left(\pd{}{t} - iq A_t \right)^2  - \frac{\partial^2}{\partial x^2}  + m^2 - i q \sigma E\right]\psi_i = 0 \,,
\end{equation}
with $\sigma = \pm 1$. We note that for $\sigma = 0$ this reduces to Eq. \eqref{kg}. Substituting the time-dependence $\exp(-i \omega t)$ into Eq. \eqref{kgdirac}, the system reduces to a Schr\"{o}dinger-like equation, corresponding to motion in the effective potential
\begin{equation}
V_\mathrm{eff} = m^2 - (\omega + q A_t)^2  - i q \sigma E \,.
\end{equation}
We require $V_\mathrm{eff}$ to be negative at $x=\pm \infty$ for asymptotic plane wave solutions to exist. Taking $A_t(-\infty)$ to be zero by convention, we find from the energetic condition  $\omega + m < q \Delta \Phi  = -q A_t(\infty)$ that $\omega + q A_t$ is positive on the asymptotic left and negative on the asymptotic right. Hence there must be a point at which $\omega + qA_t = 0$, and hence $\mathrm{Re} \, V_\mathrm{eff} = m^2 >0$. We thus see that Eq. \eqref{kgdirac} describes motion in a potential with a barrier.

It was shown by Nikishov \cite{nikishov1} that the mean number of particles produced by the field can be found by examining the scattering of plane waves off this potential barrier. In particular, if $R$ is the reflection coefficient, the mean number of particles produced is
\begin{equation}
\label{niki}
n = \pm(R - 1) \,,
\end{equation}
where the upper sign refers to bosons and the lower sign to fermions. Ordinarily $R$ would be less than unity, but in this case it can be greater, such that $n$ is always positive. 

The quantity in Eq. \eqref{niki} represents the amplification of a flux of particles incident on the electric field, a measure of the rate of stimulated emission. Indeed, the fact that the rate of \textit{spontaneous} emission is dictated by the rate of stimulated emission is reminiscent of the case of emission and absorption of photons by atoms. There is a simple argument due to Einstein that these rates are related in a direct way. Here we give an analogue schematic argument.

Denote a state with a given electric field and $n$ particle pairs by $|n\rangle$. The probability that the electric field produces another particle pair is given by $P_1 = A + nB$, where $A$ is the coefficient of spontaneous emission and $B$ the coefficient of stimulated emission, the probability being proportional to the number of pairs. The probability that the resulting state $|n+1\rangle$ decays back to the original configuration is dictated by a stimulated absorption rate $P_2 = (n+1)B'$. Since energy is conserved in Schwinger pair production, these two states should have the same energy, and so equal populations when in thermal equilibrium. Thus equilibrium demands of us that $P_1=P_2$:
\begin{equation}
A + nB = (n+1)B' \,,
\end{equation}
and for this equation to hold for any $n$, we must have $A = B = B'$. Hence the rate of stimulated emission is equal to that of spontaneous emission. We note that for bosons, this is equivalent to the well-known result in quantum mechanics that the probability for a system to decay to a state with $n$ bosons is enhanced by a factor of $n+1$ relative to the probability to decay to a state with zero bosons. Such an argument is briefly outlined for rotating black holes in \cite{microscopic}.

\subsubsection{Boundary conditions: an outgoing wave}
\label{bcs}

There is an important subtlety in the setup of this scattering problem. The boundary condition dictated by causality is that there are both ingoing and outgoing waves on one side of the potential barrier, with only an outgoing wave on the other side. 

However, what is the nature of an outgoing wave on the right-hand side (say) of the potential barrier? It is tempting to say that the solution at infinity should go as $\exp(ikx)$, as opposed to $\exp(-ikx)$. That is, would we write $\phi_L = \tilde{A} \exp(ikx) + \tilde{B} \exp(-ikx)$ and $\phi_R = \tilde{C} \exp(ikx)$. This represents a positive flux towards the right, but is not correct. The reason is that the particle on the right has energy $\omega + q A_t(\infty)$, which is \textit{negative}. For the particle to be moving in the positive direction, its group velocity must be positive. With negative energy, it is hence also necessary that the particle have negative momentum, and so the correct boundary condition to impose is that the wave on the right has the form $\exp(-ikx)$. In the above and the following, any momenta denoted by $k$ are \textit{implicitly positive}. This argument applies equally to bosons and fermions. 

\begin{figure}[h]
\centering
    \makebox[\textwidth]{\makebox[1.1\textwidth]{%
    \begin{minipage}{.55\textwidth}
        \centering
        \includegraphics[width=1\textwidth]{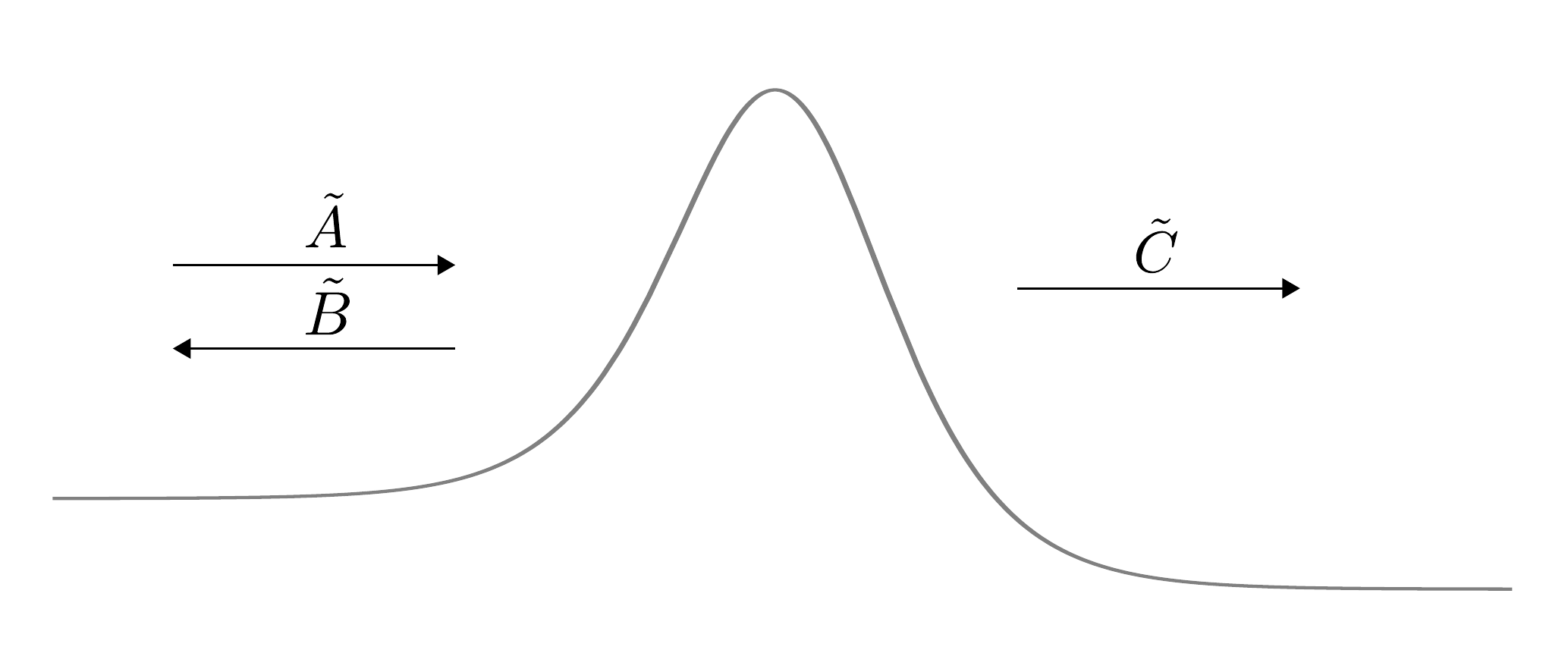}
    \end{minipage}\hfill
    \begin{minipage}{.55\textwidth}
        \centering
        \includegraphics[width=1\textwidth]{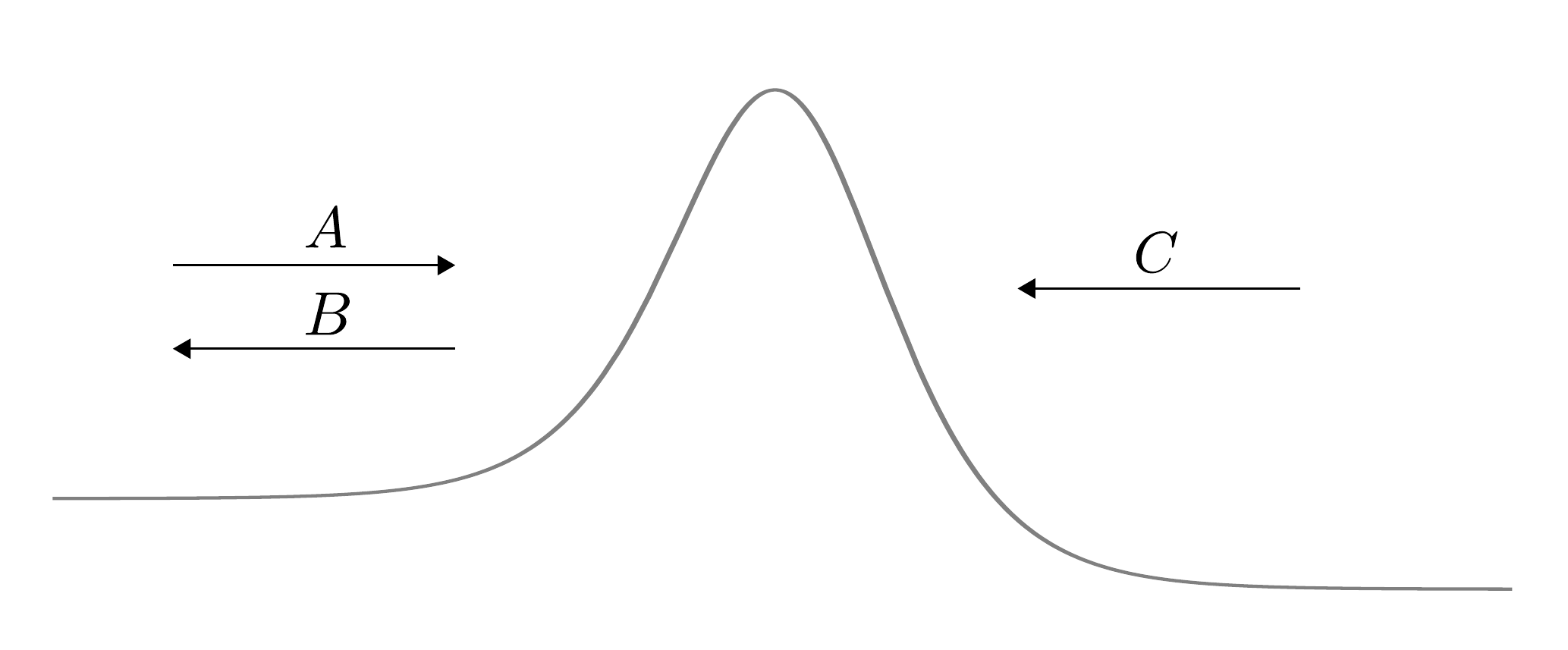}
    \end{minipage}}}
\caption{The naive (left) and physical (right) setup of fluxes.}
\label{flux}
\end{figure}
We'll define the transmission $T$ and reflection $R$ coefficients by
\begin{align}
T &= |C/A|^2 \,, \\
R &= |B/A|^2 \,,
\end{align}
and denote their naive counterparts with a tilde. We make the observation that the physical setup of fluxes is simply the time-reversal of the naive setup, provided we exchange the ingoing and reflected waves. This allows us to relate
\begin{equation}
\tilde{R} = \frac{1}{R} \,.
\end{equation}
It will transpire that with the physical setup of fluxes, the reflection coefficient will be greater than unity for bosons, and less than unity for fermions (as we might expect from the Pauli exclusion principle). With the naive setup, the reflection coefficient is instead greater than unity for fermions --- this is the result known as the Klein paradox. The resolution of the `paradox' lies in the very fact that particle-antiparticle pairs can be produced by the electric field \cite{nikishov0}, but we make pains to note that, despite the original formulation of the paradox, it is in fact \textit{bosons} which are amplified upon incidence with a strong electric field, not fermions. 

\subsubsection{Flux conservation}

To relate the reflection and transmission coefficients for the above problem, we need to understand the relevant flux conservation equations. For bosons, there is an obvious conserved current,
\begin{equation}
J_b(\phi) =  i\phi \partial_x \phi^* - i\phi^* \partial_x \phi \,,
\end{equation}
which is conserved in the sense that $\partial_x J_b = 0$. For the naive boundary conditions described above, this gives the conservation law
\begin{equation}
\label{bosecons}
k_L |\tilde{A}|^2 - k_L|\tilde{B}|^2 = k_R|\tilde{C}|^2 \,.
\end{equation}
For fermions, the fact that the effective potential in Eq. \eqref{kgdirac} is complex means this quantity is not conserved. In this case, we must instead use the conservation of the underlying Dirac current 
\begin{equation}
J_f(\psi) =  \bar{\psi} \gamma^1 \psi \,.
\end{equation}
In the asymptotic regions $x = \pm \infty$, the solutions to the Dirac equation are plane waves. Writing $\Omega = \omega + q A_t$, and labelling the components of $\psi$ by $\psi_i$, one can show that
\begin{equation}
J_f(\psi) \propto \frac{1}{\Omega}\sum_i J_b(\psi_i) \,.
\end{equation}
Since $\Omega$ is negative on the asymptotic right, we see that though a particle on the left with spatial dependence $\exp(ikx)$ corresponds to positive flux, a particle on the right with spatial dependence $\exp(ikx)$ corresponds to negative flux. For the naive boundary conditions, this conservation law becomes
\begin{equation}
\label{fermcons}
\frac{k_L}{\Omega_L}|\tilde{A}|^2 - \frac{k_L}{\Omega_L}|\tilde{B}|^2 = \frac{k_R}{\Omega_R}|\tilde{C}|^2 \,,
\end{equation}
where we note the term on the right-hand side is negative. In the simplest case that $k_L = k_R$, the two naive conservation equations Eqs. \eqref{bosecons} and \eqref{fermcons} reduce to 
\begin{align}
 |\tilde{A}|^2 - |\tilde{B}|^2 &= |\tilde{C}|^2 \,,\\
|\tilde{A}|^2 - |\tilde{B}|^2 &= -|\tilde{C}|^2 \,.
\end{align}
In terms of the physical setup of fluxes, the right-hand sides of Eqs. \eqref{bosecons} and \eqref{fermcons} obtain an additional minus sign. Dividing through by $|A|^2$, the conservation laws can be written:
\begin{align}
\label{randt}
KT \coloneqq \frac{k_R}{k_L}\,T &=  R-1  & &\mathrm{bosons}\,,\\
\label{randt2}
KT \coloneqq \left|\frac{k_R \Omega_L}{k_L \Omega_R}\right| T &= 1-R & &\mathrm{fermions} \,,
\end{align}
where we denote by $K$ the positive prefactors involving ratios of momenta. We can relate the naive and physical transmission coefficients using the relation $\tilde{R}=1/R$:
\begin{align}
\label{relations}
T &=   \frac{\tilde{T}}{1-K \tilde{T}} & &\mathrm{bosons}\,,\\
\label{relations2}
T &=  \frac{\tilde{T}}{1 +K \tilde{T}}  & &\mathrm{fermions} \,.
\end{align}
From here we find that the mean number $n$ of particles produced in the mode $\omega$, given by $n = \pm (R-1) = K T$, is
\begin{align}
n_b &= \frac{K\tilde{T}}{1-K\tilde{T}} & &\mathrm{bosons}\,, \\
n_f &= \frac{K\tilde{T}}{1+K\tilde{T}}  & &\mathrm{fermions} \,.
\end{align}

\subsubsection{A point particle perspective}
\label{ppperspective}

In \cite{kleinparadox} and \cite{remarks} an alternative argument is given for the relation between the two transmission probabilities above. An equivalent argument is also given in more detail in \cite{umetsu}. The idea is that $K\tilde{T} = P$ corresponds to the \textit{relative} probability of producing a single particle pair, in the sense that the probability of producing a state with $n$ pairs is the product of $P$ and the probability of producing a state with $n-1$ pairs. Denoting by $N$ the probability of producing no particles, we can fix $N$ by demanding that the sum of all probabilities is unity. For bosons, an arbitrary number of particles can be produced, and so
\begin{equation}
N(1 + P + P^2 + \cdots) = 1 \qquad \implies \qquad N = 1-P \,.
\end{equation}
We can then straightforwardly compute the expected number of particles produced:
\begin{equation}
\label{nb}
n_b =  N(P + 2P^2 + 3P^3 + \cdots ) = \frac{NP}{(1-P)^2} = \frac{P}{1-P} \,.
\end{equation}
For fermions, the argument is similar. The probability of producing no particles is $N$, and of producing one particle is $NP$. So for the probabilities to add to unity, we need
\begin{equation}
N = \frac{1}{1+P} \,.
\end{equation}
The expected number of particles produced is then simply
\begin{equation}
\label{nf}
n_f = NP = \frac{P}{1+P} \,.
\end{equation}
We see that these equations \eqref{nb}, \eqref{nf} are precisely the relations from the previous subsection between $\tilde{T}$ and $T$.

\pagebreak
\section{Total Rate of Radiation}
\label{total}

The phenomenon whereby particles incident on an electric field can have amplified reflected flux applies equally to the electric field outside a black hole. In this context, the amplification is known as \textit{superradiance}. See \cite{page04} for further discussion. Just as in the flat spacetime case, this gives rise to the possibility of spontaneous production of charged particle pairs outside the horizon, in addition to the thermal production of those same particles. In this section, we give formulae for the total rate of emission from a black hole, accounting for both of these processes.

The absorption cross-section $\sigma(\omega)$ in the Hawking formula Eq. \eqref{stefan} can be related by standard scattering arguments to the reflection coefficient of waves incident on the black hole thus:
\begin{align}
\label{scat}
\sigma(\omega) = \frac{\pi}{k^2}\sum_\ell (2 \ell + 1) (1 - R_\ell) \,.
\end{align}
We see that the Hawking radiation formula Eq. \eqref{stefan} contains this same $(1-R)$ factor that determines the Schwinger pair production rate Eq. \eqref{niki}. Indeed, in a more thorough derivation of Eq. \eqref{stefan} this factor need not be put in by hand. In such an analysis, the total number of particles produced can be related to the decomposition of waves in the asymptotic future into waves in the asymptotic past. This method yields both the thermal factor in the denominator, which can be attributed to the transmission of the wave through the time-dependent collapsing background, as well as the amplification factor $(1-R)$ in the numerator, which can be attributed to the transmission of the wave out through the resulting electric field, corresponding to the Schwinger process.

As well as altering the nature of the reflection coefficient $R$, the presence of an electric field also modifies the thermal factors appearing in Eq. \eqref{stefan}. The exact rate of radiation from a black hole is thus
\begin{equation}
\label{exact}
-\diff{M}{t} = \frac{g}{2 \pi^2}\int_m^\infty \mathrm{d}\omega \, \frac{\omega^2 \,k\, \sigma(\omega)}{\exp((\omega-q\Phi)/T_\mathrm{BH}) \pm 1} \,.
\end{equation}
We can then trade the reflection coefficients $R_\ell$ appearing in Eq. \eqref{scat} for transmission coefficients using Eqs. \eqref{randt} and Eq. \eqref{randt2}:
\begin{align}
\sigma(\omega) &= (\omega- q \Phi)\frac{\pi}{k^3}\sum_\ell (2 \ell + 1) T_\ell & &\text{bosons}\,, \\
\sigma(\omega) &= \frac{\pi \omega}{k^3}\sum_j (2 j + 1) T_j & &\text{fermions}\,.
\end{align}
Substituting these results into Eq. \eqref{exact}, we find that the exact rate that a black hole loses energy through emission of charged bosons is
\begin{equation}
\label{bosonexact}
\boxed{-\diff{M}{t} = \frac{g}{2 \pi}\sum_{\ell}(2\ell+1)\int_m^\infty \mathrm{d}\omega \, \frac{\omega^2}{k^2}(\omega-q\Phi)\frac{ T_\ell}{\exp((\omega-q\Phi)/T_\mathrm{BH}) - 1}\,.}
\end{equation}
Expressed in terms of the transmission factor, we can resolve a possible objection with Eq. \eqref{exact}. Note that although the denominator in the Bose-Einstein distribution gives rise to a simple pole at $\omega = q \Phi$, this is cancelled by the factor $(\omega - q \Phi)$ coming from the flux conservation equation. When the black hole is extremal, this factor will ensure that the rate of radiation at $\omega = q \Phi$ is zero, in contrast to the claims of \cite{khriplovich}, which suggests that energies close to $q \Phi$ dominate the total emission integral on account that $T_\ell$ is largest there. Likewise, this extra flux factor ensures the integrand is everywhere positive. For fermions, the exact rate of emission is
\begin{equation}
\label{fermionexact}
\boxed{-\diff{M}{t} = \frac{g}{2 \pi}\sum_{j}(2j+1)\int_m^\infty \mathrm{d}\omega \, \frac{\omega^3}{k^2}\frac{ T_j}{\exp((\omega-q\Phi)/T_\mathrm{BH}) + 1}\,.}
\end{equation}
We note that at zero temperature, the exponential factors becomes step functions that are zero for $\omega > q \Phi$, and the resulting expressions are precisely those one expects for radiation occurring solely due to the Schwinger mechanism:
\begin{align}
-\diff{M}{t} &= \frac{g}{2 \pi}\sum_{\ell}(2\ell+1)\int_m^{q \Phi} \mathrm{d}\omega \, \frac{\omega^2}{k^2}(q\Phi - \omega) T_\ell & &\text{bosons}\,,
\\
-\diff{M}{t} &= \frac{g}{2 \pi}\sum_{j}(2j+1)\int_m^{q\Phi} \mathrm{d}\omega \, \frac{\omega^3}{k^2}T_j & &\text{fermions}\,.
\end{align}
These formulae are simply the black hole analogues of Eq. \eqref{niki}, correctly accounting for the phase-space and flux factors. 

In Section \ref{combined} we will justify the formulae Eqs. \eqref{bosonexact} and \eqref{fermionexact} as arising from a combined process of tunnelling through both the horizon and the electric field.  In the rest of this section, we describe how to calculate this transmission coefficient, firstly by setting up the relevant ODE, and then by defining the appropriate boundary conditions.

\subsection{The boson equation}
The equation governing a charged bosonic particle in this background is the Klein-Gordon equation:
\begin{equation}
 -D^\mu D_\mu \phi + m^2 \phi = 0 \,.
\end{equation}
Taking our field $\phi$ to have time-dependence $\exp(-i\omega t)$ and spherical-harmonic angular-dependence, this becomes
\begin{equation}
 -\frac{f}{r^2}\diff{}{r}\left( r^2 f \diff{}{r} \right) \phi + f\left(m^2  + \frac{\ell(\ell+1)}{r^2} \right) \phi- \left(\omega - \frac{qQ}{r}\right)^2 \phi = 0 \,.
\end{equation}
To simplify this equation, we define a tortoise coordinate $r_*$ by
\begin{equation}
\label{tortoise}
\diff{r_*}{r} = 1/f \,,
\end{equation}
and rescale our field
\begin{equation}
\Psi = r \phi \,.
\end{equation}
In terms of these new quantities, our equation becomes
\begin{equation}
\label{boseeqn}
-\ddiff{\Psi}{r_*} + \left(1 - \frac{2M}{r} + \frac{Q^2}{r^2}\right)\left(m^2 +\frac{\ell(\ell+1)}{r^2} +  \frac{2M}{r^3} - \frac{2 Q^2}{r^4} \right)\Psi - \left(\omega - \frac{qQ}{r} \right)^2 \Psi = 0 \,.
\end{equation}
We have reduced the Klein-Gordon equation to a simple ODE --- indeed, as in Eq. \eqref{kgdirac}, this is a Schr\"{o}dinger-like equation, with an effective potential that has two pieces; one due to the gravitational field, and one due to the electromagnetic field. We plot the form of this potential for a typical choice of the underlying parameters in Fig. \ref{veff}.

\subsection{The fermion equation}
The equation governing a charged fermion in this background is the Dirac equation:
\begin{equation}
\slashed{D} \psi = \gamma^\mu (\nabla_\mu - iq A_\mu) \psi = \gamma^\mu (\partial_\mu + \Omega_\mu  - iq A_\mu) \psi = m \psi \,,
\end{equation}
where $\Omega_\mu$ is the spin-connection. As in Eq. \eqref{diracsq}, we can diagonalise this equation in spinor space by squaring it --- we discuss this in detail in Appendix \ref{appendixa}. Note that the equations governing the behaviour of spin-1/2 particles in a black hole background were analysed in detail by Teukolsky \cite{teuk}. However, these apply to massless particles, which is not appropriate for our purposes, and also make use of the Newman-Penrose formalism \cite{newpen}, which is more machinery than is necessary for analysing our spherically symmetric problem. After appropriate manipulations and field redefinitions, the Dirac equation can also be cast in the form of a Schr\"{o}dinger-like equation, where each component satisfies
\begin{multline}
\label{fermeqn}
-\ddiff{\Psi}{r_*} + \left(1-\frac{2M}{r}+\frac{Q^2}{r^2}\right)\left(m^2 + \frac{(j+1/2)^2}{r^2} - i \sigma \frac{q Q}{r^2}\right)\Psi\\ - \left(\omega - \frac{qQ}{r} \right)^2 \Psi+ \left(\frac{M}{r^2}-\frac{Q^2}{r^3}\right)\left(\frac{\mathrm{d}}{\mathrm{d}r_*} +  i \sigma \left( \omega -  \frac{q Q}{r}\right)\right) \Psi  = 0 \,,
\end{multline}
with $\sigma = \pm 1$ denoting the sign of the spin of the fermion and $j$, the total angular momentum quantum number, taking on the values $j = k+1/2$ for any non-negative integer $k$. 

\subsection{Boundary conditions: ingoing at the horizon}

We explained in Section \ref{hawkunch} that the appropriate boundary condition to impose in solving the transmission problem is that the wave is purely ingoing at the black hole horizon. However, we also saw in Section \ref{bcs} that there were some subtleties in defining the direction of a wave when an electric potential is present. In this section we explicitly clarify the nature of the boundary conditions.

The field equation for both bosons and fermions at the horizon is simply\footnote{This is not obvious from the fermion equation Eq. \eqref{fermeqn}, but can be seen by returning to the first-order Dirac equation. See Eq. \eqref{simpledirac} in Appendix \ref{appendixa}.}
\begin{equation}
-\ddiff{\Psi}{r_*} =(\omega - qQ/r_+)^2 \Psi \,.
\end{equation}
The general solution there is given by
\begin{equation}
\Psi = C \exp(-i \Omega r_*) + D \exp(i \Omega r_*) \,,
\end{equation}
where $\Omega = \omega - qQ/r_+$. When $\Omega$ is positive, and the emission corresponds purely to Hawking radiation, there is no difficulty: an ingoing wave corresponds to one with $D = 0$. On the other hand, when $\Omega < 0$, it is the second term which represents a wave with ingoing momentum, and so we might be tempted to claim that $C = 0$ is the appropriate boundary condition. 

However, as argued in Section \ref{bcs}, we must take care to note that since the energy of the wave is also negative, a negative momentum would give rise to a positive group velocity, i.e., an outgoing wave. The correct physical boundary condition is hence a wave with outgoing momentum. Thus, irrespective of the value of $\omega$ relative to $q \Phi$, the correct boundary condition to impose is that, at the horizon,
\begin{equation}
\Psi = C \exp (-i \Omega r_*) \,.
\end{equation}

At infinity, both field equations reduce to
\begin{equation}
-\ddiff{\Psi}{r_*} = (\omega^2-m^2) \Psi = k^2 \Psi \,,
\end{equation}
with general solution
\begin{equation}
\Psi = A \exp(-i k r_*) + B \exp(i k r_*) \,.
\end{equation}
The transmission coefficient $T$ is then defined simply by $|C/A|^2$ for bosons, and for fermions by $\sum_i |C_i|^2 / \sum_i |A_i|^2$, where the index labels the spinor component.

\section{Radiation as Tunnelling}

In this section we discuss how the emission of charged particles from charged black holes can be viewed as a tunnelling process, whereby the particle tunnels both through the horizon of the black hole (corresponding to Hawking radiation) and the electric field outside it (corresponding to Schwinger pair production). We first review the argument of Wilczek and Parikh \cite{wilczek} that Hawking radiation can be viewed as tunnelling, in the context of uncharged radiation from a Schwarzschild black hole. This picture is discussed further in  \cite{selfinteraction,secret,energycons,tunnellingmethods}, and the specific argument given below is outlined in more detail in \cite{extended}. 

For any emission process involving tunnelling, the rate of emission will have an exponential dependence of the form $\exp(-2\, \Im S) $, where $S$ is the tunnelling action. On the other hand, thermal emission of particles with frequency $\omega$ should be suppressed by an exponential of the form $\exp(-\omega/T_\mathrm{BH})$. We can hence read off the temperature at which the black hole radiates according to
\begin{equation}
T_\mathrm{BH} = \frac{\omega}{2 \,\Im S} \,.
\end{equation}
Since we wish to consider the path of a particle as it crosses the horizon, it is necessary to use a coordinate system which is regular there. We will hence study the process in Painlev\'{e}-Gullstrand coordinates $(u,r,\theta,\phi)$. The time coordinate $u$, just as for Schwarzschild time, corresponds to the time measured by a stationary observer at infinity. Ignoring angular directions, the metric is
\begin{equation}
\mathrm{d}s^2 = -\left(1-\frac{2M}{r}\right)\mathrm{d}u^2 + 2 \sqrt{\frac{2M}{r}}\, \mathrm{d}u\, \mathrm{d}r + \mathrm{d}r^2 \,.
\end{equation}
We can write the action for a particle moving freely in a curved background as
\begin{equation}
\label{action}
S = \int p_\mu \, \mathrm{d}x^\mu \qquad p_\mu = m g_{\mu \nu}\diff{x^\nu}{\tau} \,,
\end{equation}
where $\tau$ is the proper time along the worldline of the particle and $p^\mu$ coincides with its physical 4-momentum.  The radial dynamics of massive particles in Schwarzschild spacetime are determined by the equations
\begin{align}
\label{reqnm}
\left(1- \frac{2M}{r}\right)\dot{u}^2 - 2\sqrt{\frac{2M}{r}}\dot{r}\dot{u} - \dot{r}^2 = 1 \,,\\
\label{teqnm}
\left(1- \frac{2M}{r}\right)\dot{u} - \sqrt{\frac{2M}{r}}\dot{r} = \omega \,.
\end{align}
The second equation is the geodesic equation corresponding to the time-independence of the metric; in terms of the momentum defined in Eq. \eqref{action}, it can be written $p_u = -\omega$, and so $\omega$ has the interpretation of the energy of the particle as measured at infinity. Solving these equations for $\dot{u}$ and $\dot{r}$ yields
\begin{align}
\left(1-\frac{2M}{r}\right) \dot{u} &= \omega \pm \sqrt{2M/r}\sqrt{\omega^2 - 1 + 2M/r} \,, \\
\dot{r} &= \pm\sqrt{\omega^2 - 1 +2M/r} \,.
\end{align}
From here we compute the imaginary part of the action thus:
\begin{align}
\Im S &= \Im \int p_r \, \mathrm{d}r\\
&= \Im \int \left(\frac{\dot{r} + \omega\sqrt{2M/r}}{1-2M/r} + \dot{r}\right) \mathrm{d}r \\
 &= \Im \int \left(\frac{\sqrt{\omega^2 - 1 + 2M/r} + \omega\sqrt{2M/r}}{1-2M/r}\right) \mathrm{d}r \,.
\end{align}
The integrand has a pole at $r=2M$, the horizon. Choosing the prescription to integrate clockwise around this pole (into the upper-half complex-$r$ plane), we find
\begin{equation}
\Im S = 4 \pi M \omega \,.
\end{equation}
giving $T_\mathrm{BH}=1/8 \pi M$, as expected.

\subsection{Tunnelling through the horizon}
We now mirror the above line of reasoning for a charged particle moving in the background of a Reissner-Nordstr\"{o}m black hole. In Painlev\'{e}-Gullstrand coordinates, now defined by $\mathrm{d}u = \mathrm{d}t + \sqrt{1-f}/f \mathrm{d}r$, the metric takes the form
\begin{equation}
\mathrm{d}s^2 = -\left(1-\frac{2M}{r} + \frac{Q^2}{r^2}\right)\mathrm{d}u^2 + 2\sqrt{\frac{2M}{r}-\frac{Q^2}{r^2}}\, \mathrm{d}u\,\mathrm{d}r + \mathrm{d}r^2 \,.
\end{equation}
The equations of motion analogous to Eq. \eqref{reqnm} and Eq. \eqref{teqnm}  are
\begin{align}
g^{\mu \nu}p_\mu p_\nu &= -m^2 \,, \\
P_u \coloneqq p_u + qA_u &= -\omega \,,
\end{align}
where $p_\mu$ is defined as above. We can eliminate $p_u$ from these equations to determine an expression for $p_r$ (assuming purely radial motion as before):
\begin{equation}
p_r = \frac{1}{f}\sqrt{1-f}(\omega+qA_u) + \frac{1}{f}\sqrt{(\omega+qA_u)^2 - fm^2} \,.
\end{equation}
Since the action for a charged particle involves the canonical momentum $P_\mu$, and hence the gauge potential, we need to understand the behaviour of this potential at the horizon. In the usual coordinates, $A = -Q/r \, \mathrm{d}t$. Unlike the metric, however, this is legitimately singular at the horizon, on account that the form $\mathrm{d}t$ is singular there but the prefactor $-Q/r$ is well-behaved. This is not a problem however, but merely an indication that we are working in a singular gauge. We can perform a singular gauge transformation that makes $A$ smooth everywhere (for $r>0$) --- for instance, choosing the gauge function to be precisely the difference between the coordinates $u$ and $t$, we find
\begin{equation}
A = -\frac{Q}{r} \, \mathrm{d}u \,.
\end{equation}
Since $A$ is now well-defined and real everywhere, it doesn't contribute an imaginary part to the tunnelling action. We need only worry about the contribution of $p_r$. Thus
\begin{equation}
\Im S = \Im \int P_\mu \mathrm{d}x^\mu = \Im \int p_\mu \mathrm{d}x^\mu = 2\pi\, \mathrm{Res}\left(\frac{1}{f}\right)\left(\omega-qQ/r_+ \right) \,,
\end{equation}
where $\mathrm{Res}$ indicates the residue of the pole at $r=r_+$. We thus have a tunnelling probability of the form
\begin{equation}
\label{boltz}
\exp(-2\, \Im S) = \exp(-(\omega - q \Phi)/T_\mathrm{BH}) \,,
\end{equation}
where $T_\mathrm{BH}$ is precisely the temperature in Eq. \eqref{temp}:
\begin{equation}
1/T_\mathrm{BH} = 4 \pi\, \mathrm{Res}\left(\frac{1}{f}\right) \quad \implies \quad T_\mathrm{BH} = \frac{\sqrt{M^2-Q^2}}{2 \pi r_+^2} \,.
\end{equation}
The factor in Eq. \eqref{boltz} is precisely the Boltzmann factor for a particle with energy $\omega$ in an electric potential $\Phi$.

To determine the mean number of particles produced as a result of this tunnelling process, we can use the arguments of Section \ref{ppperspective} that relate tunnelling probabilities to the expected number of particles produced:
\begin{align}
n_b &= \frac{P}{1-P}  & &\text{bosons}z,,\\
n_f &= \frac{P}{1+P} & &\text{fermions}\,.
\end{align}
Treating the quantity $\exp(-2 \,\Im S)$ as the relative probability $P$ of producing a particle outside the horizon, the expected number of particles produced is
\begin{align}
n_b &= \frac{1}{\exp((\omega - q \Phi)/T_\mathrm{BH}) - 1} & &\text{bosons} \,,\\
n_f &= \frac{1}{\exp((\omega - q \Phi)/T_\mathrm{BH}) + 1} & &\text{fermions} \,.
\end{align}
These are none other than the usual Bose-Einstein and Fermi-Dirac distributions.

\subsection{Tunnelling through the electric field}
\label{tunnelelec}

We have already established that the mean number of particles emitted by the black hole depends on the transmission coefficient $T$ for fields with ingoing boundary conditions at the horizon. In general, determining this transmission coefficient requires us to numerically solve the field equations discussed in Section \ref{total} (see, for example, \cite{superradiancebrazil}). However, we can find approximate expressions for $T$ using the WKB method. A similar analysis was performed in \cite{schwingereandm} and \cite{schwingerinstantons} for the flat spacetime case, and in \cite{remarks} for charged black holes.

To the lowest level of approximation, the WKB method implies
\begin{equation}
K\tilde{T} = \exp(-2S) \,,
\end{equation}
where $S$ is defined explicitly below. In fact, as shown in \cite{froman}, a more accurate form of the transmission coefficient is given by
\begin{align}
\label{frotrans}
K\tilde{T} = \frac{1}{\exp(2S) \pm 1}  \,,
\end{align}
where the upper sign refers to bosons and the lower sign to fermions, and where the result applies to the naive transmission problem. Though the WKB analysis only applies when $S$ is large, and though these two expressions agree to first order in $\exp(-2S)$, the second expression is useful because it happens to coincide with the \textit{exact} form of the transmission probability in the flat space constant-field case --- this occurs because the relevant field equation reduces to the Schr\"{o}dinger equation for a particle moving in an (inverted) harmonic oscillator, a system for which higher-order WKB corrections vanish. 

\begin{figure}[h]
\centering
\includegraphics[width=0.8\textwidth]{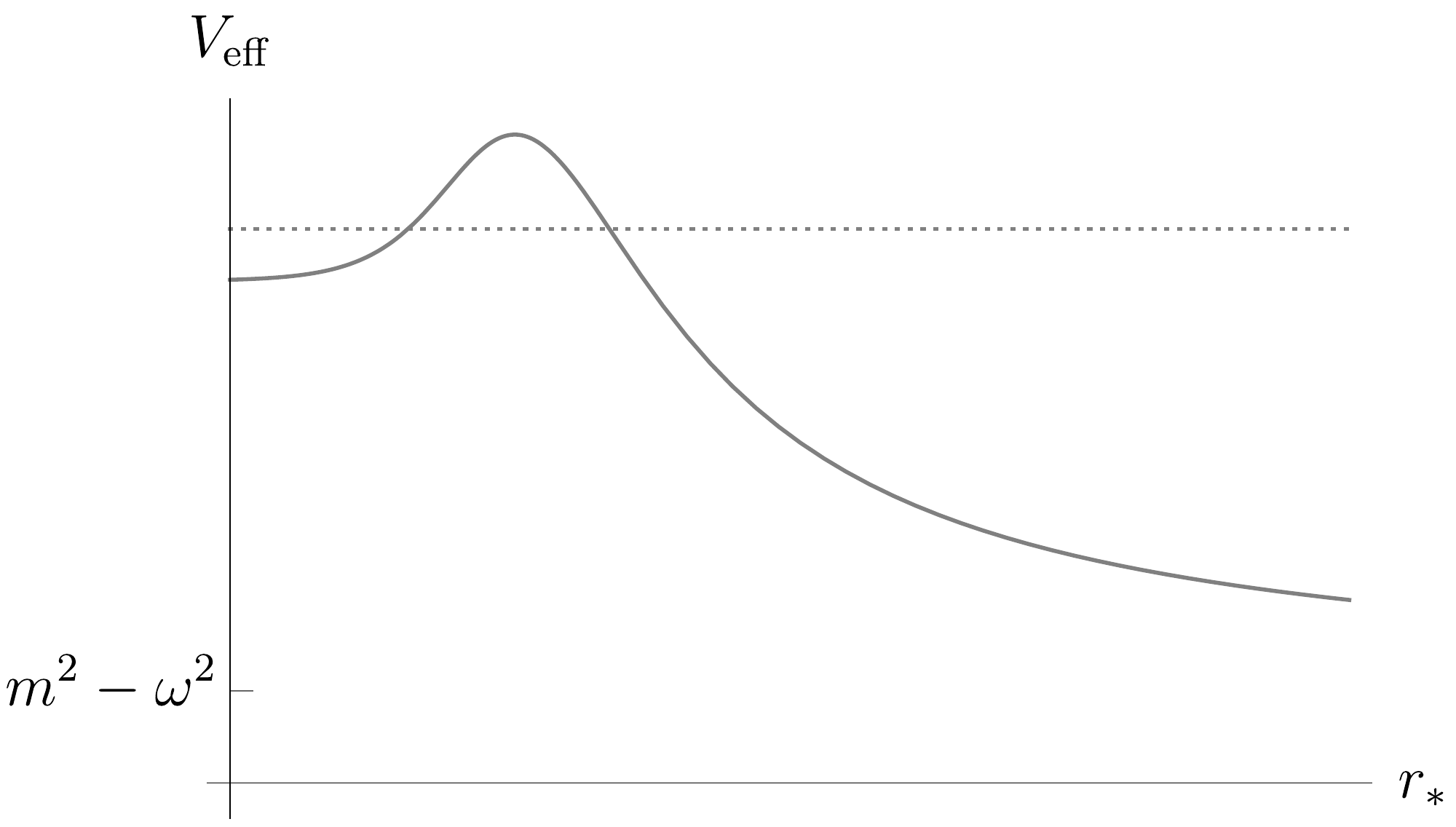}
\caption{The tunnelling barrier, as defined by the function under the square root in Eq. \eqref{largem}, for an illustrative choice of $\omega$. The dotted line represents the zero of energy --- tunnelling occurs through the region in which the potential lies above this line.}
\label{veff}
\end{figure}

Combining this more exact formula with the relations between the naive and correct forms of the transmission coefficient Eqs. \eqref{relations} and \eqref{relations2}, we find that the WKB approximation gives us the deceptively simple
\begin{eqnarray}
\label{wkbschw}
KT = \exp(-2S) \,,
\end{eqnarray}
for both bosons and fermions. The tunnelling integrals are
\begin{equation}
S = \int\sqrt{ \left(1 - \frac{2M}{r} + \frac{Q^2}{r^2}\right)\left(m^2 + \frac{\ell(\ell+1)}{r^2} +  \frac{2M}{r^3} - \frac{2 Q^2}{r^4} \right)  - \left(\omega - \frac{qQ}{r} \right)^2 } \, \mathrm{d}r_* \,,
\end{equation}
for bosons and 
\begin{equation}
S = \int\sqrt{ \left(1 - \frac{2M}{r} + \frac{Q^2}{r^2}\right)\left(m^2 + \frac{(j+1/2)^2}{r^2} \right)  - \left(\omega - \frac{qQ}{r} \right)^2 } \, \mathrm{d}r_* \,,
\end{equation}
\pagebreak

\noindent for fermions\footnote{We refer to Appendix \ref{appendixa} for a derivation of this result.}. In Fig. \ref{veff} we plot the effective potential appearing underneath the square root in these equations, in the large $m$ limit.

We note that equation Eq. \eqref{wkbschw} is not applicable when the particle momentum is in any region of space small --- in particular, it does not apply as $\omega \to q \Phi$. In that limit, $K T \to 0$ for bosons on account that the momentum factor $(\omega - q \Phi)$ goes to zero there.

\subsection{The large $m$ limit: particles}

\begin{figure}[h]
\centering
\includegraphics[width=0.8\textwidth]{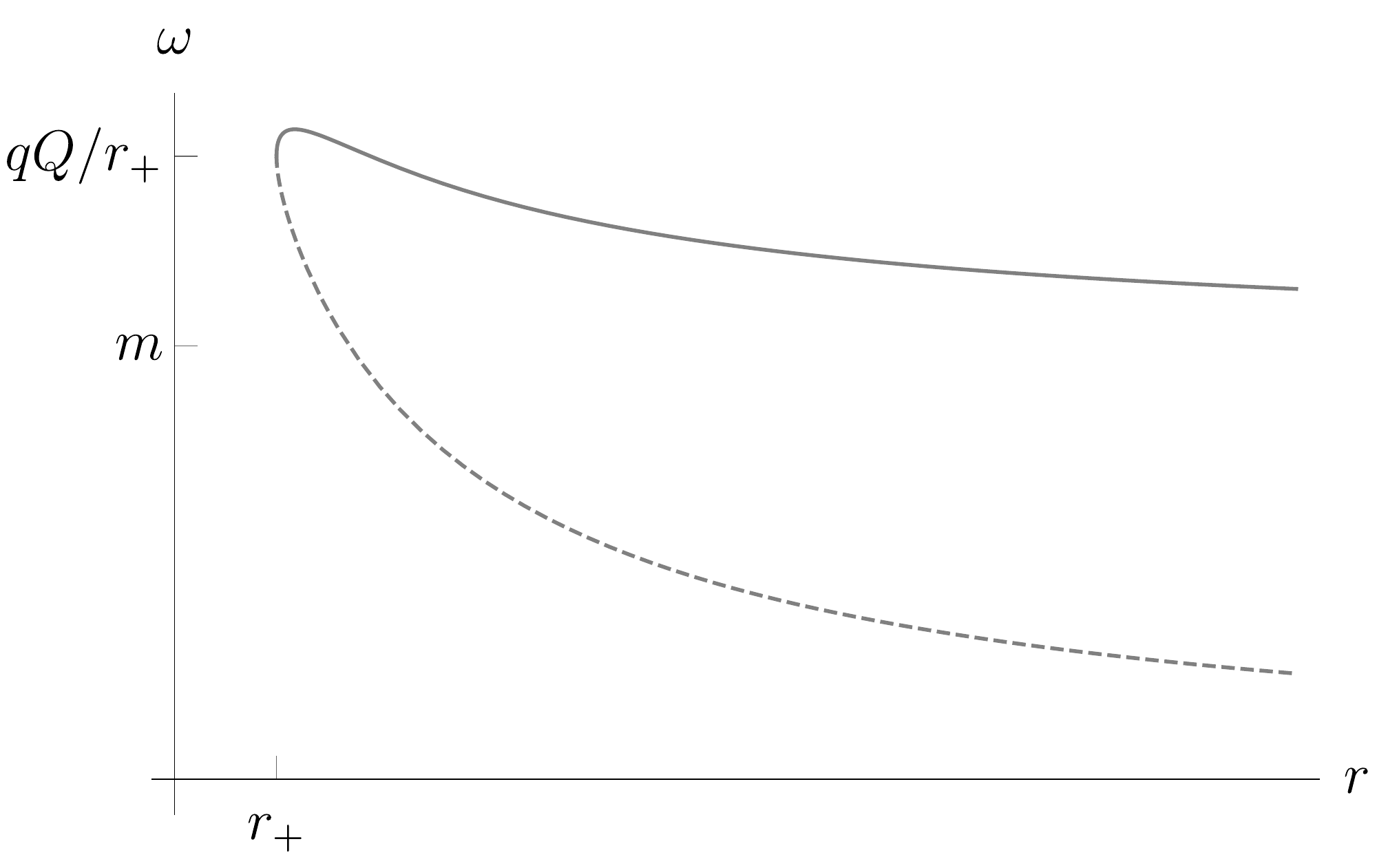}
\caption{The region in the $\omega$--$r$ plane in which there exists a tunnelling barrier, in the large $m$ limit. The dotted line corresponds to energies less than $q\Phi$ --- in this regime, there is always a barrier to emission. For a small range of energies greater than $q\Phi$, however, there also exists a barrier to tunnelling.}
\label{khrip}
\end{figure}

The tunnelling integrals discussed above are difficult to evaluate in general. However, they simplify greatly in the limit that the Compton wavelength of the particle is much less than the radius of the black hole, $m M \gg 1$. In this case, the tunnelling integral reduces both for fermions and bosons to
\begin{eqnarray}
\label{largem}
S = \int \sqrt{\left(1-\frac{2M}{r}+\frac{Q^2}{r^2}\right)m^2 - \left(\omega - \frac{qQ}{r}\right)^2} \, \mathrm{d}r_* \,.
\end{eqnarray}
In the limit that the black hole is much larger than the wavelength of the particle, we expect to be able to understand the emission from a point particle perspective, without reference to field equations. Indeed, we note that a radially moving relativistic point particle in our black hole background has dispersion relation
\begin{equation}
\label{dispersion}
g^{\mu \nu} p_\mu p_\nu = -m^2 \quad \implies \quad\frac{1}{f}\left(\omega - \frac{qQ}{r}\right)^2 - f p_r^2 = m^2 \,.
\end{equation}
The radial momentum $p_r$ is hence imaginary between the two radii $r_1, r_2$ determined solving Eq. \eqref{dispersion} with $p_r=0$. The action for a particle moving from the horizon to infinity thus acquires an imaginary part given by
\begin{equation}
\Im S = \int_{r_1}^{r_2} |p_r(r)|\,\mathrm{d}r  = \int_{r_1}^{r_2} \sqrt{\left(1-\frac{2M}{r}+\frac{Q^2}{r^2}\right)m^2 - \left(\omega - \frac{qQ}{r}\right)^2} \, \mathrm{d}r_* \,.
\end{equation}
This is precisely the integral in Eq. \eqref{largem}. We plot in Fig. \ref{khrip} the two radii $r_1, r_2$ for different choices of energy $\omega$.

The integral Eq. \eqref{largem} was analysed in \cite{khriplovich} for $\omega < qQ/r_+$. It is given by
\begin{equation}
\Im S = \frac{\pi m^2}{k(k+\omega)}\left(qQ - (\omega - k) M\right) \,,
\end{equation}
where $k = \sqrt{\omega^2-m^2}$. We can see from Fig. \ref{khrip}, however, that there is also a region of imaginary radial momentum for some energies $\omega > q Q/r_+$. This tunnelling barrier would be unimportant if the only emission process were Schwinger pair production, since particles with energies greater than $q Q/r_+$ would not be emitted. However, Hawking emission is of course possible for particles with energy greater than $q Q/r_+$, and the rate of such emission will hence be suppressed by an additional tunnelling exponential. We find that the maximum energy for which a barrier exists is
\begin{equation}
\label{omegamax}
\omega_\mathrm{max} = m^2 \sqrt{\frac{M^2 - Q^2}{Q^2 (q^2 - m^2)}} + \frac{q}{Q}\frac{q^2 Q^2 - m^2 M^2}{M(q^2 - m^2) + q\sqrt{ (q^2 - m^2) (M^2 - Q^2)}} \,,
\end{equation}
and that this occurs at a radius
\begin{equation}
r_\mathrm{max} = M \frac{\alpha^2 - Q^2}{\alpha^2 - M^2} + \alpha\sqrt{M^2 - Q^2} \frac{\sqrt{\alpha^2 - Q^2}}{\alpha^2 - M^2} \,,
\end{equation}
where $\alpha = qQ/m$. For particles with energies in the range $q Q /r_+ < \omega < \omega_\mathrm{max}$, the tunnelling integral is given by
\begin{equation}
\Im S/\pi = \frac{\omega q Q - 2\omega^2 M + m^2M}{k} + \frac{q M Q + \omega Q^2 - 2 \omega M^2}{\sqrt{M^2-Q^2}} \,.
\end{equation}
We emphasise the qualitative result that, at least in the point-particle limit, there will be an additional exponential suppression of radiation with energy less than $\omega_\mathrm{max}$, as compared to the spectrum predicted by the Bose-Einstein or Fermi-Dirac distributions.

\subsection{A combined tunnelling process}
\label{combined}

We have so far seen how both the thermal distribution factors and the transmission coefficients appearing in Eqs. \eqref{bosonexact} and \eqref{fermionexact} can be calculated by treating emission as a tunnelling process. In this final section, we use the reasoning of Section \ref{ppperspective} to provide an interpretation of the total emission rate, which involves the product of these two factors, as a combined process of tunnelling, through the horizon and subsequently the electric field.

We view emission as a two-stage process. Firstly, the particle tunnels through the horizon. For particles with $\omega > q\Phi$, the resultant particle is real (in that it has positive energy at the horizon) and can escape to infinity. For particles with negative energy at the horizon, we can view this tunnelling process as readjusting the particle numbers in the Dirac sea outside the horizon. Particles in this Dirac sea can then tunnel through the electric field to infinity. We will frame the analysis in terms of density matrices for notational ease, although these need only be thought of as describing probability distributions.

For fermions this picture is clearer. The tunnelling action $P_H = \exp(-(\omega-q\Phi)/T_\mathrm{BH})$ describes the probability of producing one particle in the Dirac sea \textit{relative} to producing none. Denoting particles in the Dirac sea with a bar, this yields the density matrix
\begin{equation}
\rho_H = N_H |\bar{0}\rangle \langle \bar{0}| + N_HP_H|\bar{1}\rangle \langle \bar{1}|\,,
\end{equation}
where $N_H = 1/(1+P_H)$ ensures the probabilities sum to unity. Only the state in which there is a particle in the Dirac sea can tunnel to become a real particle at infinity. As before, the relative probability of this process is $P_S = K \tilde{T}$. Hence the density matrix at infinity is
\begin{equation}
\rho_\infty = N_H |\bar{0}\rangle \langle \bar{0}| + N_H  P_H N_S|\bar{1}\rangle \langle \bar{1}| + N_H P_H  N_S P_S|1\rangle \langle 1| \,,
\end{equation}
where  $N_S = 1/(1+P_S)$. The mean number of particles detected is then simply
\begin{equation}
n_f = \mathrm{tr}(n \rho_\infty) = N_H N_S P_H P_S = \frac{1}{1+\exp((\omega-q\Phi)/T_\mathrm{BH})}\frac{K\tilde{T}}{1+K\tilde{T}} \,,
\end{equation}
where $n$ is the number operator that counts particles at infinity. This is precisely the factor appearing in Eq. \eqref{fermionexact}. 

For bosons, the picture is somewhat murkier, on account that there is no simple description of the bosonic vacuum in terms of a Dirac sea of negative energy particles. However, such interpretations have been proposed, as in \cite{diracsea3}, for instance. In such a picture, the negative energy bosonic states can contain any \textit{negative} number of particles, with the vacuum corresponding to the state with negative one particles per mode. Denoting a state with $-n$ particles by $|\bar{n}\rangle$, for continuity with the fermionic notation, we can view the process of tunnelling through the horizon as repopulating these negative energy states thus:
\begin{equation}
\label{rhoh}
\rho_H = N_H \Big( |\bar{1}\rangle \langle \bar{1}| + \frac{1}{P_H} |\bar{2}\rangle \langle \bar{2}| + \frac{1}{P_H^2} |\bar{3}\rangle \langle \bar{3}| + \cdots \Big) \,,
\end{equation}
where $N_H = 1-1/P_H$. When thermal effects are absent, the system is described by the state $|\bar{1}\rangle$, and this state can lead to production of an arbitrary number of charged particles at infinity. In particular, the probability for $n$ particles to be produced is given by a geometric distribution with relative probability $P_S$:
\begin{equation}
|\bar{1}\rangle\langle\bar{1}| \to N_S \Big( |0\rangle \langle 0| + P_S|1\rangle \langle 1| + P_S^2 |2\rangle \langle 2|+ \cdots \Big) \,,
\end{equation}
where $N_S = 1-P_S$. We next ask what the resulting state at infinity is for the doubly-occupied state $|\bar{2}\rangle$, after tunnelling through the electric field. If each `particle' in this state is independent of the other, we expect the resulting probability distribution to be the sum of two independent geometric distributions with the same mean. Likewise, we expect the state $|\bar{n}\rangle$ to tunnel to a state described by $n$ independent geometric distributions. The sum of independent and identical geometric distributions is described by the negative binomial distribution. In particular, we have
\begin{equation}
|\bar{n}\rangle \langle \bar{n}| \to (1-P_S)^n \sum_{k=0}^\infty \binom{n+k-1}{k}  P_S^k \,|k\rangle\langle k| \,.
\end{equation}
Combining this tunnelling process with the horizon tunnelling process in Eq. \eqref{rhoh} gives the density matrix at infinity in the bosonic case:
\begin{equation}
\label{rhoinf}
\rho_\infty = N_H\sum_{n=1}^\infty P_H^{1-n} (1-P_S)^n  \sum_{k=0}^\infty \binom{n+k-1}{k}  P_S^k \,|k\rangle\langle k| \,.
\end{equation}
From here, as before, we can calculate the expected number of particles detected at infinity. One finds
\begin{equation}
n_b = \mathrm{tr}(n \rho_\infty) = \frac{1}{1-\exp((\omega-q\Phi)/T_\mathrm{BH})}\frac{K\tilde{T}}{1-K\tilde{T}} \,,
\end{equation}
which is precisely the factor appearing in Eq. \eqref{bosonexact}.

The probability distribution given in Eq. \eqref{rhoinf} thus reproduces the mean number of particles produced by the black hole. We note, however, that this picture of tunnelling also provides distinct, testable predictions for the variance, skewness and higher moments of the distribution of number of particles produced.

\section{Discussion}
 
We have given exact formulae for the rate of emission of charged particles from charged black holes, taking care to define the differential equations that govern this process, the appropriate boundary conditions for those differential equations, and to specify precisely the phase-space and flux factors appropriate to massive particles for which $\omega \neq k$. Concrete expressions for the transmission coefficients have been given in the point-particle limit, and in particular we have found that for particles with energy below $\omega_\mathrm{max}$, given in Eq. \eqref{omegamax}, emission rates from black holes will be exponentially suppressed relative to energies above it. In addition to justifying the formulae for the average number of particles received at infinity, we have also given the expected probability distribution for the number of these particles, in a given mode. This provides new predictions for, e.g., the uncertainty in the number of particles received.

We have restricted attention to non-rotating black holes in this work. We note, however, that rotating black holes have many similar properties to charged black holes --- in particular, there is a region of energies for which flux directed onto the black hole is reflected with larger amplitude. In this context, such superradiance goes by the name of the Penrose process \cite{penrose}. Quantum mechanically, we thus expect a rotating black hole to spontaneously emit spinning particles (in such a way as to reduce its angular momentum), although it is not clear what the appropriate flat spacetime analogue of this process would be. It would be interesting in future work to examine how such spontaneous emission could also be viewed as a tunnelling process, this time through a vacuum spacetime, but one with a more non-trivial gravitational field structure.

\section*{Acknowledgements}
I am grateful to John March-Russell for many useful discussions, and to the Science and Technology Facilities Council (STFC) for financial support.

\appendix

\section{The Dirac Equation in a BH Background}
\label{appendixa}

Using Latin indices $a,b$ to denote a normalised basis aligned with the $(t,r,\theta,\phi)$ coordinate system, we define our gamma matrices by
\begin{equation}
\{ \gamma^a, \gamma^b \} = 2 \eta^{a b} \,.
\end{equation}
We can choose, for instance,
\begin{equation}
\gamma^0 = i\left(\begin{array}{C C C C}
0 & 0 & 1 & 0\\
0 & 0 & 0 & 1 \\
1 & 0 & 0 & 0 \\
0 & 1 & 0 & 0
\end{array}\right)\qquad \gamma^1 = i\left(\begin{array}{C C C C}
0 & 0 & 1 & 0\\
0 & 0 & 0 & -1 \\
-1 & 0 & 0 & 0 \\
0 & 1 & 0 & 0
\end{array}\right) \,,
\end{equation}
to ensure that $\sigma^{01}$ is diagonal:\begin{equation}
-i\sigma^{01} = \frac{1}{2} \gamma^0 \gamma^1  = \frac{1}{2}\mathrm{diag}(1,-1,-1,1)
\,.\end{equation}
The Dirac equation in curved spacetime, in the presence of an electromagnetic field, is
\begin{equation}
\slashed{D} \psi = \gamma^\mu (\nabla_\mu - iq A_\mu) \psi = \gamma^\mu (\partial_\mu + \Omega_\mu  - iq A_\mu) \psi = m \psi \,,
\end{equation}
where $\Omega_\mu$ is the spin connection. Further discussion can be found in \cite{curvedfermions}. We will first rewrite this equation in the case $A_\mu = 0$, restoring the electric field later. In terms of the tortoise coordinate defined by Eq. \eqref{tortoise}, and writing $\partial_*$ to mean $\partial/\partial r_*$, the Dirac equation can be written explicitly as
\begin{equation}
\label{curvdirac}
\gamma^\mu \partial_\mu \psi + \frac{1}{2}\left(\frac{\gamma^1}{\sqrt{f}}\left(\frac{2}{r} - \frac{3M}{r^2} + \frac{Q^2}{r^3}\right) + \gamma^2\frac{\cot \theta}{r} \right) \psi = m \psi \,,
\end{equation}
with
\begin{equation}
\gamma^\mu \partial_\mu = \frac{\gamma^0}{\sqrt{f}}\partial_t + \frac{\gamma^1}{\sqrt{f}}\partial_* + \frac{\gamma^2}{r}\partial_\theta + \frac{\gamma^3}{r \sin \theta}\partial_\phi  \,.
\end{equation}
Naively squaring this equation yields the second-order form
\begin{equation}
\left(\nabla^2_S + 2 \Omega \cdot \partial + \Omega^2 + \nabla_\mu \Omega^\mu  \right) \psi = m^2 \psi \,,
\end{equation}
where $\nabla^2_S$ is the scalar Laplacian. Explicitly, the operator on the left-hand side is
\begin{multline}
\frac{1}{f}\left(-\partial_t^2 + \partial_*^2 \right) + \frac{2}{r} \partial_* + \frac{L_S^2}{r^2} -\frac{1}{4f}\left(\frac{M}{r^2}-\frac{Q^2}{r^3}\right)^2 -\frac{f}{2r^2} - \frac{\cot^2 \theta}{4r^2} \\
-\frac{2}{f}\left(\frac{M}{r^2}-\frac{Q^2}{r^3}\right) i \sigma^{01} \partial_t + \frac{\sqrt{f}}{r^2} i \sigma^{12}(\cot \theta + 2 \partial_\theta) + \frac{2 \cot \theta}{r^2 \sin \theta} i \sigma^{23} \partial_\phi + \frac{2 \sqrt{f}}{r^2 \sin \theta} i \sigma^{13} \partial_\phi \,,\end{multline}
where $L_S^2$ is the usual angular momentum operator in quantum mechanics. This unwieldy equation is not only non-diagonal in spinor space, unlike its flat space counterpart, but it does not reduce to the wave equation at the horizon or spatial infinity. We instead first perform the following field redefinition (see \cite{neutrinos}, for instance)
\begin{equation}
\Psi = r f^{1/4} \sqrt{\sin \theta} \,\psi \,,
\end{equation}
which vastly simplifies the first-order equation Eq. \eqref{curvdirac}:
\begin{equation}
\label{simpledirac}
\gamma^\mu \partial_\mu \Psi = m \Psi \,.
\end{equation}
One can always choose the spinors to be eigenfunctions of the angular operator
\begin{equation}
L \Psi \coloneqq \left(\partial_\theta + \frac{\partial_\phi}{\sin \theta}\right) \Psi =  -i \lambda \gamma^0 \gamma^1 \Psi \,,
\end{equation}
where the eigenvalue $\lambda$ satisfies $\lambda^2=(j+1/2)^2$. Since the particle has spin half, the total angular momentum $j$ can be any half-integer, and so $\lambda^2$ can be any positive square integer. Multiplying Eq. \eqref{simpledirac} through by $\sqrt{f}$ are squaring gives
\begin{equation}
\left(-\partial_t^2 + \partial_*^2\right) \Psi = f\left(m^2 + \frac{\lambda^2}{r^2}\right)\Psi + \gamma^1 \partial_* \left(m\sqrt{f} + \frac{i\lambda \sqrt{f}}{r} \gamma^0 \gamma^1 \right) \Psi \,.
\end{equation}
We can further simplify this equation by substituting in the first-order form:
\begin{equation}
\left(-\partial_t^2 + \partial_*^2\right) \Psi_i = f\left(m^2 +\frac{\lambda^2}{r^2}\right) \Psi_i + \left(\frac{M}{r^2}-\frac{Q^2}{r^3}\right) (\partial_* - \sigma \partial_t) \Psi_i + \frac{i \lambda f^{3/2}}{r^2} (\gamma^0 \Psi)_i \,,
\end{equation}
where $\sigma = \pm 1$ is the sign of the spin of the fermion. The final term in this equation is the only one which is not diagonal in spinor space. Since we will be analysing this equation using the WKB method, and since this term is sub-leading to both the second derivative terms, which go as $\omega^2$, as well as the first derivative terms, which go as $\omega$, we will henceforth ignore this term.

Reintroducing the electromagnetic potential yields a broadly similar equation. There is an additional spin-field coupling (morally the $\sigma^{\mu \nu}F_{\mu \nu}$ term in Eq. \eqref{diracsq}):
\begin{multline}
-\left(\partial_t + \frac{iq Q}{r} \right)^2 \Psi_i + \partial_*^2 \Psi_i = f\left(m^2 +\frac{(j+1/2)^2}{r^2}\right)\Psi_i \\+ \left(\frac{M}{r^2}-\frac{Q^2}{r^3}\right) \left(\partial_* - \sigma \left(\partial_t +  \frac{iq Q}{r}\right)\right) \Psi_i - i\sigma f \frac{qQ}{r^2}\Psi_i \,.
\end{multline}
As we discuss in Appendix \ref{appendixb}, to leading order in the WKB approximation, we can simplify this equation to
\begin{equation}
\left(\omega- \frac{q Q}{r} \right)^2 \Psi_i + \ddiff{\Psi_i}{r_*} = f\left(m^2 +\frac{(j+1/2)^2}{r^2}\right) \Psi_i \,,
\end{equation}
where we've substituted the time-dependence $\exp(-i\omega t)$.

For completeness, we give the relation between the tortoise coordinate $r_*$ and the original radial coordinate $r$. For non-extremal black holes we have
\begin{equation}
r_* = r - \frac{2M^2-Q^2}{2\sqrt{M^2-Q^2}}\,\ln\left(\frac{r - M + \sqrt{M^2 - Q^2}}{r - M - \sqrt{M^2 - Q^2}}\right) + M \ln \left(\frac{r^2- 2Mr + Q^2}{4M^2}\right) \,,
\end{equation}
whilst for extremal black holes we have instead
\begin{equation}
r_* = r + 2 M \ln\left(\frac{r-M}{M}\right) - \frac{M^2}{r-M} \,.
\end{equation}

\section{The WKB Solution}
\label{appendixb}

Here we provide a quick review of the WKB solution of an ordinary differential equation. See also \cite{froman} and \cite{bender} for more information. Suppose we have an equation of the form
\begin{equation}
\ddiff{y}{x} + U(x)\diff{y}{x} + V(x) = 0 \,,
\end{equation}
where $V(x)$ is in some sense large, say of order $\mu^2$. Then we try a solution of the form
\begin{equation}
y = \exp(iW(x)) \,,
\end{equation}
and expand $W = \mu W_0 + W_1 + \cdots$ and $V = \mu^2 V_0 + \mu V_1 + \cdots$ and $U = U_0 + \cdots$ in powers of $\mu$. The leading and next-to-leading order equations read:
\begin{align}
-\mu^2(W_0')^2 + \mu^2 V_0 &= 0 \,,\\
i\mu W_0'' - 2 \mu W_0' W_1' + i\mu U_0 W_0' +\mu V_1 &= 0 \,.
\end{align} 
The zeroth-order equation is solved by
\begin{equation}
W_0 = \pm\int \sqrt{V_0} \, \mathrm{d}x \coloneqq \int k \, \mathrm{d}x \,,
\end{equation}
and hence the first-order equation becomes
\begin{equation}
\label{firstorder}
2 W_1' = i\frac{k'}{k} + i U_0 + \frac{V_1}{k} \,.
\end{equation}
If $V_1$ is imaginary and $U_0$ and $k$ are real, this will mean $W_1$ is pure imaginary, and so the first-order equation dictates how the amplitude of the wave varies with position. For illustration, we can consider the Dirac equation above with $\lambda = q = 0$ for simplicity. Then
\begin{align}
V_0 &= \omega^2 - f m^2   \,,\\
V_1 &= -\frac{i \omega \sigma}{2f} \diff{f}{r_*} \,,\\
U_0 &= -\frac{1}{2f} \diff{f}{r_*}\,.
\end{align}
Consider $\sigma = -1$ --- we know from Eq. \eqref{simpledirac} that a wave with this spin must be outgoing at the horizon, which corresponds to taking $k > 0$. In this case we can in fact solve Eq. \eqref{firstorder} for $W_1$:
\begin{equation}
2W_1 = i \ln k - \frac{i}{2} \int \mathrm{d}r_* \, \frac{1}{f}\diff{f}{r_*}\left(1-\frac{\omega}{k}\right) \,.
\end{equation}
Rewriting $k$ in terms of $f(r_*)$ leaves us with an integral we can perform analytically:
\begin{align}
2W_1 &= i \ln k - \frac{i}{2} \int \mathrm{d}f \, \frac{1}{f}\left(1-\frac{1}{\sqrt{1 - fm^2/\omega^2}}\right)\\
&=  i \ln k - i \ln\left(1 + \sqrt{1-f m^2/\omega^2}\right) \,.
\end{align}
Our WKB solution thus becomes
\begin{equation}
\psi(r_*) \propto \left(\frac{\omega + k(r_*)}{k(r_*)}\right)^{1/2}\exp\left(i \int^{r_*} \mathrm{d}r_* \, \sqrt{\omega^2 -  f(r_*)m^2} \right) \,.
\end{equation}
We note that the additional complications introduced by the presence of the terms $V_1$ and $U_0$ have unimportant quantitative consequences --- since the wavenumber $k$ lies in the range $0 < k(r_*) < \omega$, the numerator of the amplitude can vary by at most a factor of $\sqrt{2}$ over the domain of interest. We can hence find a good approximation to the transmission amplitude by retaining only zeroth-order terms in the WKB expansion.

\bibliography{schw}
\bibliographystyle{JHEP}

\end{document}